\documentclass[twocolumn,letterpaper]{article}

\usepackage[pagenumbers]{cvpr} 

\usepackage{multirow}
\usepackage{graphicx}
\usepackage{amsmath}
\usepackage{amssymb}
\usepackage{booktabs}
\usepackage{amsfonts,amssymb}
\newcommand*{\method}{BoostMIS}

\usepackage{bbold}
\usepackage[pagebackref,breaklinks,colorlinks]{hyperref}

\usepackage[capitalize]{cleveref}
\crefname{section}{Sec.}{Secs.}
\Crefname{section}{Section}{Sections}
\Crefname{table}{Table}{Tables}
\crefname{table}{Tab.}{Tabs.}

\def\cvprPaperID{6449
} 
\def\confName{CVPR}
\def\confYear{2022}

\begin{document}

\title{\method{}: Boosting Medical Image Semi-supervised Learning  \\ with Adaptive  Pseudo Labeling and Informative Active Annotation }

\author{Wenqiao Zhang$~\textsuperscript{\rm 1}$ \and Lei Zhu$~\textsuperscript{\rm 1}$  \and  James Hallinan$~\textsuperscript{\rm 2}$ \and   Andrew Makmur$~\textsuperscript{\rm 2}$   \and  Shengyu Zhang$~\textsuperscript{\rm 3}$ \and   Qingpeng Cai$~\textsuperscript{\rm 1}$  Beng Chin Ooi$~\textsuperscript{\rm 1}$ \\
\small{$~\textsuperscript{\rm 1}$ National University of Singapore, Singapore}, 
\small{$~\textsuperscript{\rm 2}$ National University Hospital, Singapore}, 
\small{$~\textsuperscript{\rm 3}$ Zhejiang University, China} 
\\{\tt\small wenqiao@nus.edu.sg, e0203764@u.nus.edu, $\{$james$\_$hallinan,andrew$\_$makmur$\}$@nuhs.edu.sg,}
\\{\tt\small sy$\_$zhang@zju.edu.cn, $\{$qingpeng,ooibc$\}$@comp.nus.edu.sg }
}
\maketitle
\begin{abstract}

In this paper, we propose a novel semi-supervised learning (SSL) framework named BoostMIS that combines adaptive pseudo labeling and informative active annotation to unleash the potential of medical image SSL models:
(1) BoostMIS can adaptively leverage the cluster assumption and consistency regularization of the unlabeled data according to the current learning status.  This strategy can adaptively generate one-hot ``hard'' labels converted from task model predictions for better task model training. (2) For the unselected unlabeled images with low confidence,  we introduce an Active learning (AL) algorithm to find the informative samples as the annotation candidates by exploiting virtual adversarial perturbation and model's density-aware entropy.
These informative candidates are subsequently fed into the next training cycle for better SSL label propagation.  Notably, the adaptive pseudo-labeling and informative active annotation form a learning closed-loop that are mutually collaborative to boost medical image SSL.  To verify the effectiveness of the proposed method, we collected a metastatic epidural spinal cord compression (MESCC) dataset that aims to optimize MESCC diagnosis and classification for improved specialist referral and treatment. We conducted an extensive experimental study of BoostMIS on MESCC  dataset. The experimental results verify our framework's effectiveness  with a significant improvement over various state-of-the-art methods.
\vspace{-0.3cm}
\end{abstract}

\begin{figure}[t]
\includegraphics[width=0.5\textwidth]{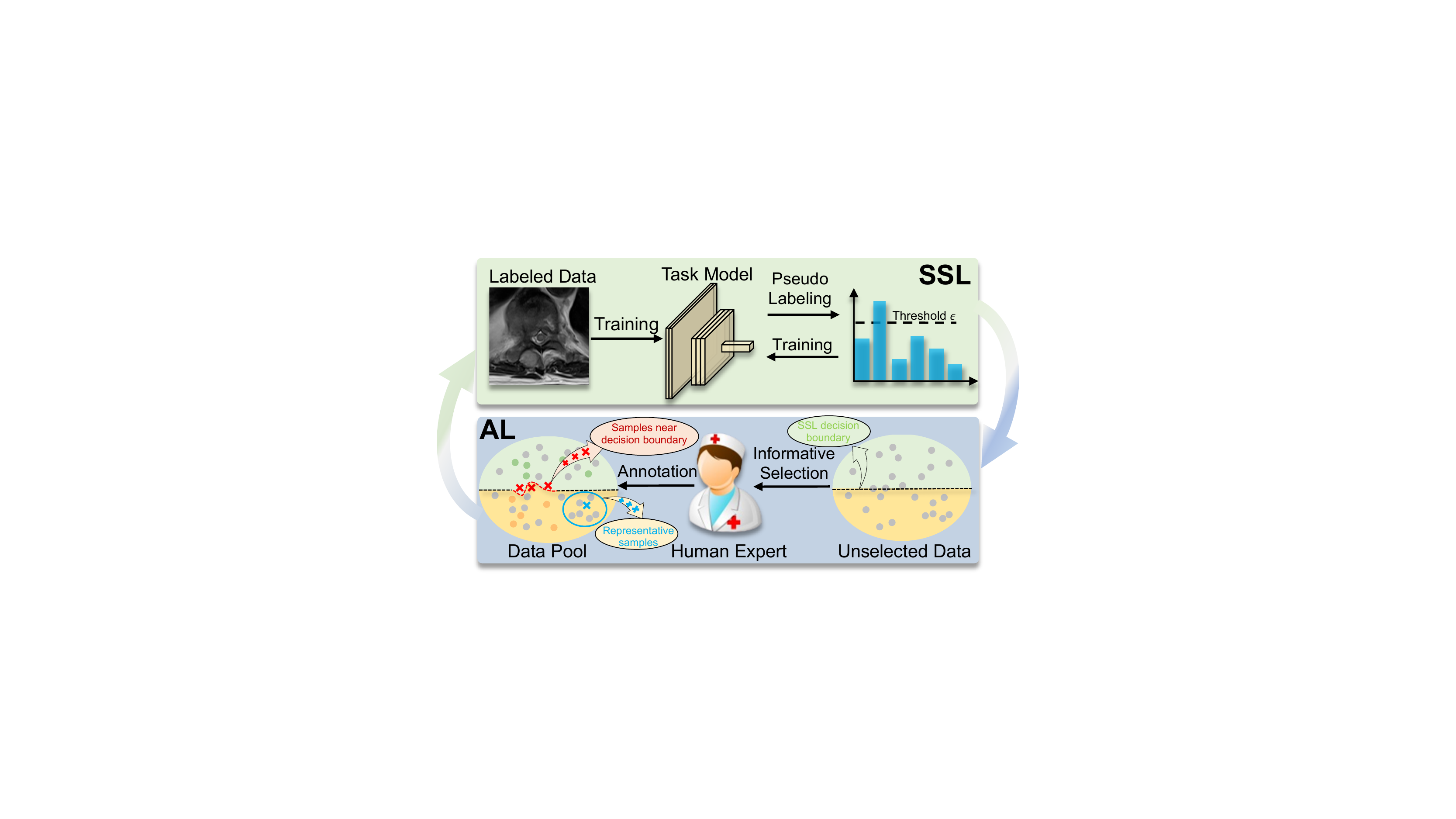}
\centering\caption{\textbf{An  example  of  how does AL facilitates medical image SSL.} Orange and green dots represent two different classes in the data pool. Grey dots represent the unlabeled data. Red and blue marks indicate the annotation candidates, which can smooth the decision boundary and propagate representative label information to unlabeled data, respectively.}
\label{fig1}
\vspace{-0.3cm}
\end{figure}
\section{Introduction}
In recent years, the development of deep learning brings prosperity to the field of computer vision~\cite{li2019walking,zhang2021tell,li2022compositional,zhang2019frame,yang2021multiple,li2021adaptive,li2022compositional,zhang2020relational,li2020multi,Li2022End2End} and applied in industrial application~\cite{li2020unsupervised,kakani2020critical,villalba2019deep,xu2021computer,zhang2021magic,li2020ib}. For instance, a significant trend in healthcare is to leverage the sizable well-annotated dataset using deep learning technology~\cite{shen2017deep,wang2018interactive,zhang2019medical,zhou2018unet++,chen2021transunet,hallinan2021deep} to achieve automated medical image analysis. 
However, annotating such a medical dataset is an expert-oriented, expensive, and time-consuming task. The situation is further exacerbated by large numbers of new raw medical images without labels. These are generated on a daily basis in clinical practice and there is an increasing demand for the exploitation of such data. Therefore, semi-supervised learning (SSL) that goes beyond traditional supervised learning by exploiting a limited amount of labeled data together with a huge amount of raw unlabeled data is gaining traction.
Among the existing SSL methods, pseudo-labeling~\cite{lee2013pseudo} is a specific variant where model predictions are converted to pseudo-labels, which is often used along with confidence-based thresholding that retains unlabeled examples only when the classifier is sufficiently confident. Some of them~\cite{sohn2020fixmatch,berthelot2019mixmatch,xie2019unsupervised,kurakin2020remixmatch}, have achieved great success on image classification datasets such as CIFAR-10/100~\cite{krizhevsky2009learning}, demonstrating the potential value of utilizing unlabeled data.

%
Unfortunately, these natural image-based pseudo-labeling SSL methods may not satisfactorily resolve the medical imaging problems, which can be summarized as two key aspects: (1) \textbf{Poor Data Utilization}.  Medical images (e.g.,  CT, MRI) are highly similar at the pixel level, making them hard to classify, even by human beings. That is, the pseudo-labeling may only produce a few pseudo-labels with high confidence above a fixed threshold for unlabeled medical images, especially at the early stage of the training process. It suffers from the poor utilization of data problem that is a considerable amount of unlabeled data are ignored. 
(2) \textbf{Missing Informative Sample}. The ignored unlabeled data with prediction confidence below the pre-defined threshold may have informative data (e.g., samples near the boundary of clusters, representative samples in the unlabeled distribution space) that can further improve the model's performance.
Based on the aforementioned insights, a meaningful optimization goal of Medical Image SSL is to explore an effective learning approach to tap onto unlabeled medical data deeply.

Inspired by another alternative to leverage the unlabeled data, that is active learning (AL)~\cite{settles2009active}, which aims to select the most informative samples to maximize the model performance with minimal labeling cost. AL seems to be a tempting approach to deal with the aforementioned problems in pseudo-labeling SSL methods. As illustrated in Figure~\ref{fig1}, SSL already results in the embodiment of knowledge from selected unlabeled data with high confidence by pseudo-labeling. The judicious AL selection can reflect the value of additionally informative samples in unselected data on top of such embodied knowledge.  Those informative cases can assist the SSL model to propagate extra valuable knowledge to unlabeled data, thereby improving the unlabeled data utilization for better SSL.
In fact, AL and SSL are naturally related with respect to their common goal, i.e., utilizing unlabeled samples. 
From the perspective of machine learning, utilizing the correlation and similarity among related learning methods can be regarded as a form of inductive transfer. It can introduce the \emph{inductive bias}~\cite{baxter2000model} to make the combined learning method prefer the correct hypothesis, thereby improving the performance.


Overall,  we propose a novel framework \method{} that aims to boost medical image SSL, which comprises the following parts: (1) \emph{Medical Image Task Model}. The task model (medical  image  classification  in this  paper) is first trained through weakly-augmented medical images with supervised labels.  (2) \emph{Consistency-based Adaptive Label Propagator}. This module propagates the label information for unlabeled data using both pseudo-labeling and consistency regularization. Since the learning ability and performance of the models are different in each training stage, we define a dynamically adaptive threshold based on the current learning status to produce pseudo-labels for better unlabeled data utilization. 
Then the consistency regularization forces the model to produce the same prediction of weakly-augmented and strongly-augmented data as the regularization criterion for better generalization capacity.
(3) \emph{Adversarial Unstability Selector}. To boost SSL by AL, we introduce the virtual adversarial perturbation to choose the unstable samples that lying on the clusters’ boundaries as annotation candidates. Specifically,  the SSL model would be weaker or even inconsistent on the clusters’ boundaries, the adversarial unstability selector could identify samples near the boundaries by measuring the inconsistency between samples and the corresponding virtual adversarial examples. (4) \emph{Balanced Uncertainty Selector}. To further identify the informative cases in the unlabeled pool,  we use the density-aware entropy of the SSL model to evenly select the samples with high uncertainty in each predicted class as the complementary set to balance the subsequent training . The union of adversarial unstability samples and balanced uncertainty samples will be the final annotation candidates to expand the labeled pool. In summary, our proposed four modules work in a loop and make progress together to boost the medical image SSL.



To summarize, the major contributions of our paper are as follows:
\vspace{-0.1cm}
\begin{itemize}
\item  To the best of our knowledge, we are the first to incorporate AL into SSL to unleash the potential of unlabeled data for better medical image analysis.

\item We propose adaptive pseudo-labeling and informative active annotation that reasonably leverage the unlabeled medical images and form a closed-loop structure to boost the medical image SSL.

\item We collected a metastatic epidural spinal cord compression (MESCC) dataset for method development and extensive evaluation, which aims to optimize MESCC  diagnosis and classification for improved specialist referral and treatment.

\item The consistent superiority of the proposed \method{} is demonstrated on the MESCC  dataset  that outperforms existing SSL methods by a large margin. 

 \end{itemize}

\section{Related Work}
\noindent\textbf{Semi-supervised Learning.}
Semi-supervised learning (SSL) is concerned with learning from both unlabeled and labeled samples, where typical methods ranging from entropy minimization~\cite{grandvalet2005semi,huang2010semi,vu2019advent}, pseudo-labeling (also called self-training)~\cite{arazo2020pseudo,ding2019feature,lee2013pseudo,xie2020self} and consistency regularization~\cite{verma2021interpolation,xie2019unsupervised,bachman2014learning,dai2017good} to improve model's performance. Recent work FixMatch~\cite{sohn2020fixmatch} that proposes to combine pseudo-labeling and consistency regularization simultaneously has shown promising performance in semi-supervised tasks. FixMatch~\cite{sohn2020fixmatch} achieves SoTA results by combining these techniques with weak and strong data augmentations and uses cross-entropy loss as the regularization criterion. However, FixMatch suffers from the poor utilization of data problem that is it ignores a considerable amount of unlabeled data with confidence below the labeling threshold. Therefore, an important goal of Medical Image SSL is to utilize the unlabeled data appropriately. \\
\noindent\textbf{Semi-supervised Learning in Medical Image Analysis.}
Due to the difficulty of data annotation, SSL is widely used in medical imaging processing, e.g., medical image  detection~\cite{wang2020focalmix,zhou2021ssmd}, classification~\cite{gyawali2020semi,shang2019leveraging,mahapatra2021medical} and segmentation~\cite{nie2018asdnet,zhou2019collaborative}. ~\cite{wang2020focalmix}  proposes an SSL method for 3D medical image detection that uses a generic generalization of the focal loss to regularize the SSL training. ~\cite{zhou2019collaborative} improves the performance of disease grading and lesion segmentation by SSL. ~\cite{nie2018asdnet} uses an attention-based SSL learning method to boost the performance of medical image segmentation. However, these previous works are limited to SSL, which may fail to mine the potential of unlabeled data deeply. 

\noindent\textbf{Semi-supervised Active Learning.} Recently, few works are proposed that combine SSL and AL. ~\cite{sinha2019variational,ebrahimi2019uncertainty} generally adopt VAE-GAN structure in the SSL manner to learn latent representations of labeled and unlabeled samples through a min-max game, and then conduct AL-based on learned semantic distribution. The latest works ~\cite{song2019combining,gao2020consistency,guo2021semi} make the combination based on prediction inconsistency given a set of data augmentations. However, all of these methods are designed for natural image classification tasks, while this paper focuses on the complicated medical image SSL task.


\begin{figure*}[t]
\includegraphics[width=0.92\textwidth]{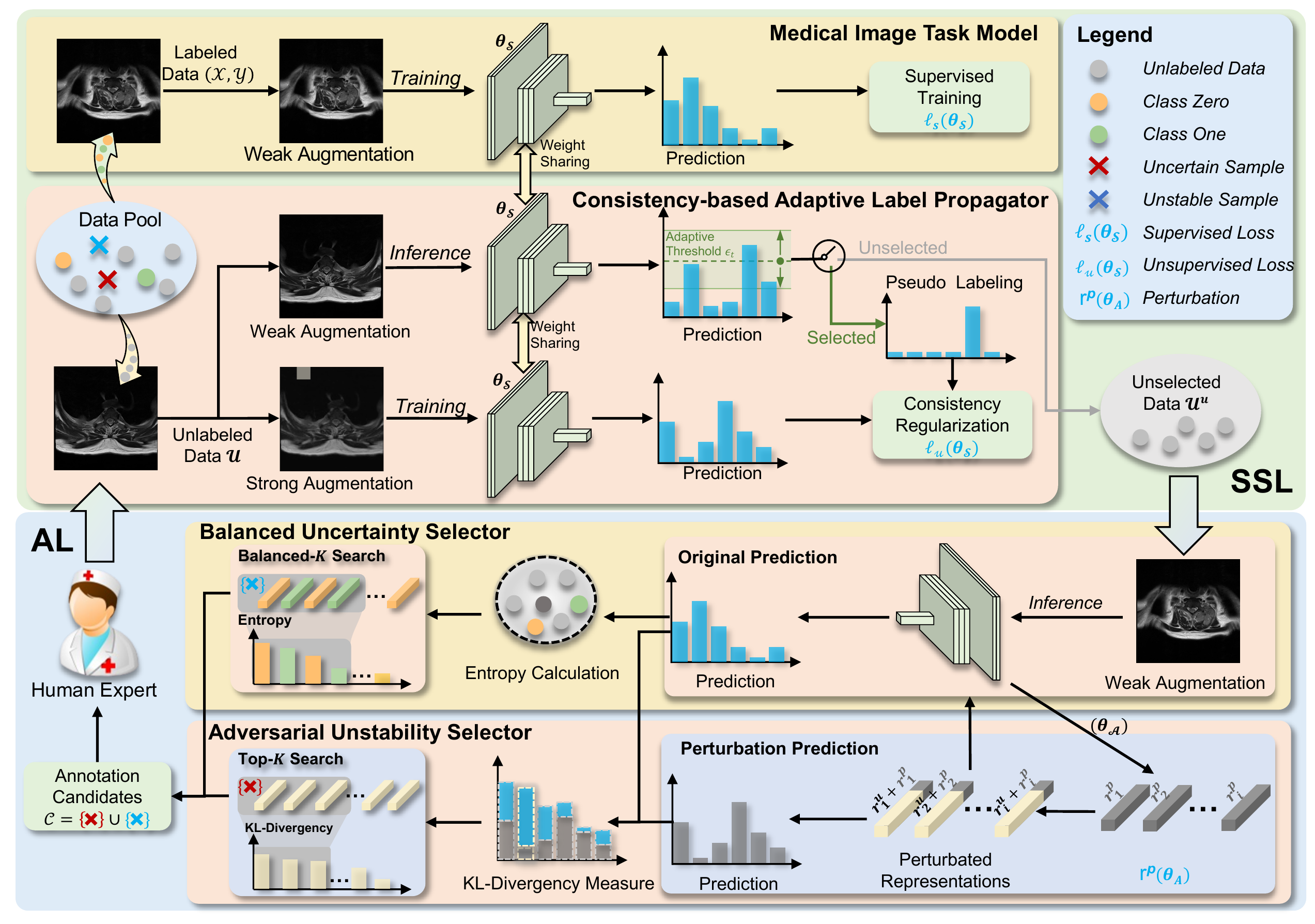}
\vspace{-0.2cm}
\centering\caption{  \textbf{Overview of  \method{}.} It consists of four modules: (a) \emph{Medical Image Task Model} trains the basic task model with weakly-augmented labeled data by supervised learning.  (b)~\emph{Consistency-based Adaptive Label Propagator} propagates the label information to unlabeled samples by pseudo-labeling with dynamically adjusted threshold and augmentation based consistency regularization. (c)~\emph{Adversarial Unstability Selector}
estimates KL-Divergence of unselected data and their corresponding virtual adversarial examples with perturbation to select the annotation candidates. (d)~\emph{Balanced Uncertainty Selector} evenly selects the samples with high uncertainty as the complementary annotation set to balance the subsequent SSL training.}

\label{fig2}
\vspace{-0.2cm}
\end{figure*}

\section{Method}
This section describes the proposed medical image semi-supervised (SSL) framework \method{}. We will describe each module and introduce the training strategy.
\subsection{Problem Formulation}
Before presenting our method, we first introduce some basic notions and terminologies. We consider a labeled pool of training samples $\mathcal{X}$ with the corresponding labels $\mathcal{Y}$, and an unlabeled pool of training samples $\mathcal{U}$. The goal is to use all paired samples ($\mathcal{X,Y}$) and unlabeled samples $\mathcal{U}$ to train the proposed medical image SSL framework \method{}. As shown in Figure~\ref{fig2}, we use the following training pipeline $M(.;\theta)$  to show how the \method{} works: 
\begin{equation}
\begin{aligned}
\underbrace{M((\mathcal{X},\mathcal{Y},\mathcal{U});(\theta_{\mathcal{S}},\theta_{\mathcal{A}}))}_{\rm{\method{}}}&=  \underbrace{T((\mathcal{Y},\mathcal{P}^s)|(\mathcal{X},\mathcal{U}^s));\theta_{\mathcal{S}})}_{\rm{SSL\ Training}} \\& \leftrightarrow
\underbrace{S((\mathcal{C}|\mathcal{U}^u);(\theta_{\mathcal{S}},\theta_{\mathcal{A}}))}_{\rm{AL\ Annotation}}
 \label{adv}
\end{aligned}
\end{equation}
$T(\cdot; \theta)$ is the task model (medical image classification in this paper) 
with SSL and $S(\cdot; \theta)$ is the annotation candidate selection through active learning (AL).  \textbf{In SSL aspect},  $T(\cdot; \theta)$ is trained by labeled samples $\mathcal{X}$ with weak augmentation in \emph{Medical Image Task Model} and propagates the label information to unlabeled samples $\mathcal{U}$ in \emph{Consistency-based Adaptive Label Propagator}.  The unlabeled samples with their posterior probability above the adaptive threshold are selected as training examples $\mathcal{U}^s$ with pseudo labels  $\mathcal{P}^s$, which combine $\mathcal{X}$ to train $T(\cdot; \theta)$ with augmentation-aware consistency regularization.  \textbf{In AL aspect}, once the SSL training process is finished,  we feed unselected samples $\mathcal{U}^u$ with perturbation to $T(\cdot; \theta)$ to generate the adversarial samples in \emph{Adversarial Unstability Selector}.  After that, we select out top-$K$ samples with the biggest KL-divergence between unselected samples $\mathcal{U}^u$ and their adversarial samples. To further identify the informative cases in the unlabeled pool,  we use the entropy of $T(\cdot; \theta)$ to evenly select the top-$K$ samples with high uncertainty in each predicted class as the complementary set to balance the subsequent training. The final candidates $\mathcal{C}$ composed of parallelly selected $N_A$  samples ($\leq 2K$ due to the intersection of candidates) are provided to human experts for annotation.  As a consequence, the sizes of the labeled pool and unlabeled pool will be updated to perform better SSL. The loop will be repeated until the task model's performance meets requirements or the annotation budget has run out.


\subsection{Medical Image Task Model}\label{sec:STM}
For task model development, we train a convolutional prototypical network~\cite{yang2018robust} with Resnet50 ~\cite{he2016deep} as its backbone.  Given labeled data $\mathcal{X}=\{\textbf{x}_i|_{i=1}^{N_l}\}$ with weak augmentation (e.g., using only flip-and-shift data augmentation) and their corresponding labels $\mathcal{Y}=\{\textbf{y}_i|_{i=1}^{N_l}\}$, $N_l$ is the number labeled samples. The task model is trained via the mapping function $F$ : $\mathcal{X} \rightarrow\mathcal{Y}$, with the cross-entropy loss $\ell_s$ minimized through supervised label information.
\vspace{-0.15cm}

\begin{equation}
     \ell_s(\theta_{\mathcal{S}}) = \frac{1}{N_l}\sum_{i=1}^{N_l}D_{ce}(\textbf{y}_{i}, P_m ({{\textbf{p}_i}|A_w(\textbf{x}_i)}))
\end{equation}
where $P_m(\cdot|\cdot)$ represents the posterior probability distribution of the task model. $D_{ce}(\cdot,\cdot)$ denote the cross-entropy between two probability distributions.  $\textbf{p}_i$ is the predicted label. $A_w(.)$ is a weak augmentation function.

\subsection{ Consistency-based Adaptive Label Propagator}\label{sec:TMST}
Inspired by FixMatch~\cite{sohn2020fixmatch}, SSL label propagator in \method{} is a combination of two approaches: pseudo-labeling and consistency regularization.
The pseudo-label of an unlabeled weakly-augmented image is computed based on the task model's prediction when its confidence is above the defined threshold, and the pseudo-label is enforced against the task model's output for a strongly-augmented image. 
However, the learning ability and performance of the models are different in each training step. Thus, we introduce adaptive threshold (\textbf{AS}) $\epsilon_{t}$ that can be dynamically adjusted at different learning statuses to utilize unlabeled data better.  Formally speaking, in each iterative training step $t$, the AS $\epsilon_{t}$ for the class with fewer samples is defined as bellow:
\vspace{-0.1cm}
\begin{equation} \!\!\!\epsilon_{t} = 
\begin{cases} 
\alpha \cdot Min\{1, \frac{Count_{\epsilon_{t}}}{Count_{\epsilon_{t-1}}}\}+\frac{\beta \cdot N_A}{2K} , \!\! & \mbox{if } t< T_{max}\\
\alpha +  \beta , & \mbox{otherwise}
\end{cases} 
\label{computeA}
\end{equation}
where coefficients $\alpha$ and $\beta$ are pre-defined thresholds. $T_{max}$ is a defined value of the iterative learning step. $N_A$ is the number of active learning selected annotation candidates. $Count_{\epsilon_{t}}$ is a counting function to estimate the learning status at time step $t$. Given $\mathcal{U}=\{\textbf{u}_i|_{i=1}^{N_u}\}$ ($N_u$ is the number of unlabeled data), $Count_{\epsilon_{t}}$ computed as follows: 
\vspace{-0.1cm}
\begin{equation} 
\vspace{-0.1cm}
Count_{\epsilon_{t}}= \sum_{i =1}^{N_u}\mathbb{1} ( P_m(\textbf{p}_i|A_w(\textbf{u}_i))>\alpha +  \beta)
\label{count}
\end{equation}
$Count_{\epsilon_{t}}$ records the number of unlabeled data $\mathcal{U}$ whose prediction scores are above the high confidence threshold $\alpha + \beta$. 
We argue that the learning effect of the SSL model can be reflected by the number of samples whose predictions are above the threshold. The AS $\epsilon_{t}$ can be dynamically adjusted based on the current learning status and amount of AL informative selection to encourage the  better utilization of unlabeled data (prediction of the class with fewer samples ) until the iteration step exceeds $T_{max}$. 

Thus, we feed the unlabeled data set $\mathcal{U}$ with weak augmentation into the task model. The predictions of selected samples $\mathcal{U}^s=\{\textbf{u}_i^s|_{i=1}^{N_u^s}\}$ are converted to one-hot pseudo-labels $\mathcal{P}^s=\{\textbf{p}_i^s|_{i=1}^{N_u^s}\}$  when their predicted confidences are above the adaptive threshold $\epsilon_{t}$, where ${N_u^s}$ denote the number of SSL selected unlabeled samples. Then we introduce the augmentation-aware consistency regularization, which computes the model’s prediction for a strong augmentation of the same image. The model is trained to enforce its prediction on the strongly-augmented image to match the pseudo-label via a cross-entropy loss. 
\begin{equation}
\vspace{-0.1cm}
\!\!\!\!\ell_u(\theta_{\mathcal{S}}) \!\!=\!\!\frac{\mu}{N_u^s}\!\sum_{i=1}^{N_u^s} D_{ce}( P_m(\textbf{p}^s_i|A_w(\textbf{u}_i^s)),P_m(\textbf{p}_i^s|A_s(\textbf{u}_i^s)))\!\!
\end{equation}
where $\mu$ is a fixed scalar hyperparameter denoting the relative weight of the unlabeled loss, $A_s(.)$ represents the strong augmentation function RandAugment~\cite{cubuk2020randaugment}.


\subsection{Adversarial Unstability Selector}\label{sec:var}
Above SSL model label propagator can deliver the label information from labeled data to pseudo-labeled samples. However, there are many informative cases among the unselected samples. These samples fail to acquire enough label information, and the task model has a vague understanding of the local data distribution of these samples. 
In this sense, selecting informative samples in unselected data for further annotation is more valuable, which can help the SSL model smooth the decision boundary and propagate the representative label information to unlabeled data.  We divide the unselected samples $\mathcal{U}^u=\{\textbf{u}_i^u|_{i=1}^{N_u^u}\}$ (${N_u^u}$ is the number of unselected samples) into two categories: \emph{unstable} and \emph{uncertain}. In this section, we aim to find the unstable samples by introducing the adversarial unstability selector (\textbf{AUS}).

The AUS  estimates the stability of the model's prediction on samples by calculating the inconsistency between prediction on samples and corresponding virtual adversarial samples.
Specifically, given an unselected sample $\textbf{u}^u_i$ ($i \leq N_u^u$), its representation $\textbf{r}^u_i$ is extracted from the last layer of task model and prediction is ${{\textbf{p}}}^u_i$.  We feed $\textbf{r}^u_i$ and ${\textbf{p}}^u_i$ simultaneously into the generator to get virtual  adversarial perturbation $\textbf{r}^{p}_i$. After that, the perturbed representation ${\bar{\textbf{r}}}^u_i = \textbf{r}^u_i + \textbf{r}^{p}_i$ is fed into the task model to get its perturbed prediction ${\bar{\textbf{p}}}^u_i$.  The perturbation is formulated as:
\begin{equation}
\textbf{r}^{p}_i \!\!= \mathop{\arg \max}_{\Delta r,||\Delta r||\leq \tau}\!\!\! Div(P_m(\textbf{p}^u_i\arrowvert \textbf{r}^{u}_i),\! P_m({\bar{\textbf{p}}}^u_i \arrowvert \textbf{r}^{u}_i + \Delta r)) \!
 \label{radv}
\end{equation}
where $\tau$ a hyper-parameter that indicates the maximum perturbation step, $\Delta r$ is a randomly sampled vector,  $Div(\cdot,\cdot)$ is a non-negative function measuring the divergence between two distributions, and we use KL divergence in practice. 

The evaluation of $\textbf{r}^{p}_i$ cannot be performed using Eq.\eqref{radv} because the gradient of $Div(\cdot,\cdot)$ with respect to $r$ is always $0$ at $r = 0$. To solve this problem, \cite{miyato2018virtual} proposed to approximate $\textbf{r}^{p}_i$ using the second order Taylor expansion and solve the $\textbf{r}^{p}_i$ via the power iteration method.
Specifically, we can approximate $\textbf{r}^{p}_i$ by repeatedly applying the following update $n_t$ times ($n_t$ is a hyper-parameter):
\begin{equation}
    \!\!\textbf{r}^{p}_i\!\! \leftarrow \tau \overline{\nabla_{\Delta r}
    Div(P_m(\textbf{p}^u_i|\textbf{r}_i^u), P_m(\bar{\textbf{p}}_i^u|\textbf{r}_i^u + \Delta r))} \!\! \label{adv}
\end{equation}
the computation of $\nabla_{\Delta r}Div$ can be performed with one iteration of backpropagation for the neural network. Once the $\textbf{r}^{p}_i$ is solved, we can estimate the variance $Var(\cdot;\cdot)$ of unlabeled sample $\textbf{r}^u_i$ by measuring the KL-divergence:
\begin{equation}
    Var(\textbf{r}^u_i;\theta_A) = Div(P_m(\textbf{p}_i^u|\textbf{r}^u_i), P_m(\bar{\textbf{p}}^u_i|\textbf{r}^u + \textbf{r}^{p}_i))
\end{equation}

Finally, AUS selects top-$K$ samples with the largest variance from unstable samples as an initial recall set of the AL annotation candidates. These unstable samples often lie on the clusters' boundary that can smooth the decision boundary for the SSL model to output more correct predictions.

\subsection{Balanced Uncertainty Selector}\label{sec:bl}
Besides unstable samples, there are abundant uncertain samples among unselected data that still hold low prediction confidence are also informative to the task model. 
To further find the uncertain samples to boost the SSL model,  we introduce the balanced uncertainty selector (BUS), which evenly chooses the unselected samples with high uncertainty in each predicted class.
These selected samples can significantly reduce model uncertainty and balance the subsequent SSL. In practice, we use the task model's entropy to estimate the uncertainty of samples, and then evenly select top-${K}$ samples with the largest entropy for human annotation. The entropy $Ent'(\cdot;\cdot)$ for the unlabeled sample $\textbf{u}_i^u$ can be calculated with the following formulation:
\vspace{-0.1cm}
\begin{equation}
     \!\!\!\!\!\!Ent'(\textbf{u}_i^u;\theta_S) \!=\!\! \sum_{c \in C}  P_m(\textbf{p}^c_i|A_w(\textbf{u}_i^u)) log P_m(\textbf{p}^c_i|A_w(\textbf{u}_i^u))\!\!\!\! \label{entropy}
     \vspace{-0.1cm}
\end{equation}
where $C$ is the set of possible classes in the dataset. $\textbf{p}^c_i$ indicates the predicted class of  $\textbf{u}_i^u$ belonging to the c-th class.

The above entropy-based formula only estimates information certainty of each sample and fails to take the distribution relations among samples into account. As a result, the metric may run into the risk of selecting some outliers or unrepresentative samples in the distribution space. To alleviate this issue, we re-weight the uncertainty metric with a representativeness factor and explicitly consider the data distribution. We denote this density-aware uncertainty as:
\begin{equation}
\vspace{-0.2cm}
Ent(\textbf{u}_i^u;\theta_S) = Ent'(\textbf{u}_i^u;\theta_S) (\frac{1}{M}\sum_{j=1}^M Sim(\textbf{u}_i^u,\textbf{u}_j^u))
\end{equation}
where  $Sim(\cdot,\cdot)$ estimates the cosine similarity of $\textbf{r}^u$ and its nearest $M$ samples in the distribution space. Each class samples $\lfloor \frac{K}{N_c} \rfloor$ images as uncertain annotation candidates. 

Finally,  the union of unstable and uncertain samples composes the final annotation candidates to expand the labeled pool, which brings informative cases to improve the subsequent SSL.
\vspace{-0.1cm}


\subsection{Training Algorithm}
Algorithm 1 (in the appendix) presents the pseudocode of our \method{}: (1) On the one hand, the label propagator can propagate the supervised label information to unlabeled samples by adaptive pseudo-labeling and augmentation-aware consistency regularization. This training strategy can mix up the pseudo-labeled samples that provide extra explicit training signals and initial labeled samples to improve the task model's performance. 
(2) On the other hand, the adversarial unstability selector and the balanced uncertainty selector let the oracle annotate the samples with the largest inconsistency and the highest uncertainty, which could assist the SSL model to include the informative samples for better label propagation. In summary, the proposed \method{} lets SSL and AL models work collaboratively and form a closed-loop naturally to boost the medical image SSL.

\section{Experiments}
We evaluate the effectiveness of our proposed \method{} in two datasets that are the metastatic epidural spinal cord compression (MESCC) dataset, followed by a discussion of  \method{}’s property with controlled studies.
\subsection{Dataset and Setting}
\noindent$\textbf{Dataset.}$  We collected a metastatic epidural spinal cord compression (MESCC) dataset that aims to aid MESCC diagnosis and classification. 
It consists of 7,295 medical images collected from adult patients ($\geq$ 18 years) with a random selection of studies across different MRI scanners (GE and Siemens 1.5 and 3.0T platforms). 
When drawing each bounding box,  the annotating radiologist employed the Bilsky classification~\cite{bilsky2010reliability} that consists of patients with six types of grades that are b0, b1a, b1b, b1c, b2, and b3. In general, we also can classify these grades into low-grade (i.e., b0, b1a, and b1b) and high-grade (i.e., b1c, b2, and b3) Bilsky MESCC. Specifically, patients with low-grade are amenable to radiotherapy, while the patients with high-grade are more likely to require surgical decompression.  
The dataset is randomly split into 70\% (5207) / 15\% (1011) / 15\% (1077) for the training/validation/test sets, respectively.  More details of the MESCC  dataset can be seen in the appendix.

\noindent$\textbf{Implementation Details.}$  
We employ the Wide ResNet-50~\cite{he2016deep} as the backbone model to conduct the medical image classification task.
To train the AL-based SSL model with a balanced initialization, we set up the initial labeled pool (10\% data) by uniformly sampling in each class. The initial labeled pool is a subset of the training set (30\% data) for pure SSL models and other data are  randomly sampled.
We train \method{} with a standard stochastic gradient descent (SGD)~\cite{bottou2010large} optimizer in all experiments. Besides, we use an identical set of hyperparameters ($\mu$=1, $Mo$=0.9, $\alpha$=0.9, $\beta$=0.05, $T_{max}$=50,000, $\tau$=1, $B$=64, $AC$=30, $IP$=10\%, $SS$=30\%)\footnote{$B$ and $Mo$ refers to batch size and  momentum in SGD optimizer. $IP$=10\% and $SS$=30\% are the sampling ratio of the initial AL labeled pool and SSL training set. $AC$ indicates the max AL training cycle. } across all datasets.

\noindent \textbf{Metrics.}  To evaluate the medical image classification task, we employ the following metrics: Accuracy ({ACC}), Macro Precision ({MP}),  Macro F1 score ({MF1})~\cite{chinchor1993muc},  Macro recall ({MRC})~\cite{chinchor1993muc} and Error Rate ({ER}).

\noindent\textbf{Comparison of Methods.}\quad 
For quantifying the efficacy of the proposed framework, we use several baselines for performance comparison according to different aspects. 
On the SSL aspect, we consider three SSL baselines: \textbf{P-Labeling}~\cite{lee2013pseudo}, \textbf{MixMatch}~\cite{berthelot2019mixmatch},  \textbf{FixMatch}~\cite{sohn2020fixmatch}. On the AL aspect, we choose the following recent methods as baselines: \textbf{R-Labeling}~\cite{figueroa2012active},   \textbf{DBAL}~\cite{gal2017deep},  \textbf{VAAL}~\cite{sinha2019variational} and  \textbf{CSAL}~\cite{gao2020consistency}. More details of baselines are in the appendix.


Further, we also investigate effectiveness of each individual component in \method{} via evaluating the following variants: (1) \textbf{ \method{} (-AS)} uses a fixed confidence threshold ($\alpha+\beta$) to propagate the label information to unlabeled data. (2) \textbf{ \method{} (-BUS)} only considers the inconsistency in AL annotation to choose the unselected medical images.
(3) \textbf{ \method{} (-VAR)} selects the AL annotation candidates without Adversarial Unstability Selector. (4) \textbf{SSL+RS} is a simplified version of \method{}, which consists of the SSL model in \method{} and applies a random sampling strategy.

\begin{table*}
\small
\caption{\textbf{Performance comparison on the MESCC dataset.}  Superscript $^{\dagger}$ indicates that the model only employs the SSL algorithm.
AL$^{*}$ indicates that the SSL model (FixMatch) uses the corresponding AL annotation strategy. We report the accuracy of all these models when the percentage of labeled pool reaches 15\%/20\%/25\%/30\%. A larger score indicates better performance, and the top two scores of accuracy are in bold. Acronym notations of each model can be found in Section 4.1.}
\centering
\begin{tabular}[width=1\textwidth]{l|c|cccc|ccc}
\toprule[1.5pt]
\multicolumn{1}{c|}{\multirow{2}{*}{\textbf{Methods}}} & \multicolumn{1}{c|}{\multirow{2}{*}{\textbf{AL$^{*}$}}} &\multicolumn{4}{c|}{MESCC   two-grading}                                                  & \multicolumn{3}{c}{MESCC   six-grading}

\\ \cmidrule(l){3-6} \cmidrule(l){7-9}
\multicolumn{1}{c|}{} & & \textbf{15\% Labels}  & \textbf{20\% Labels}  & \textbf{25\% Labels}     & \textbf{30\% Labels}    & \textbf{20\% Labels}  & \textbf{25\% Labels}         & \textbf{30\% Labels}   \\ \cmidrule(l){3-6} \cmidrule(l){1-2} \cmidrule(l){7-9}
\textbf{P-Labeling}$^{\dagger}$~\cite{lee2013pseudo} &                           & 78.64$\pm$ 5.20    & 81.34$\pm$ 4.64    & 82.54$\pm$ 5.85   & 85.61$\pm$ 4.74    & 32.43$\pm$ 5.39    & 37.88$\pm$ 3.53  &  43.16$\pm$ 3.81   \\     
\textbf{MixMatch}$^{\dagger}$~\cite{berthelot2019mixmatch}  &
& 85.42$\pm$ 2.69    & 86.44$\pm$ 2.88    & 88.77$\pm$ 1.39   & 90.81$\pm$ 1.58    & 43.18$\pm$ 4.18    & 46.98$\pm$ 1.67  &  52.83$\pm$ 0.93   \\  
\textbf{FixMatch}$^{\dagger}$~\cite{sohn2020fixmatch}  &                           
& 85.52$\pm$ 2.04    & 87.37$\pm$ 3.06    & 90.34$\pm$ 1.49   & 91.09$\pm$ 0.30  & 41.41$\pm$ 4.27    & 48.65$\pm$ 1.76  &  54.32$\pm$ 1.02     \\ 
 \midrule[1pt]
\textbf{R-Labeling}~\cite{figueroa2012active}  &  \checkmark                  
& 83.66$\pm$ 5.29    & 84.12$\pm$ 5.91    & 87.47$\pm$ 3.34   & 88.67$\pm$ 3.53     & 38.07$\pm$ 6.69    & 42.34$\pm$ 5.76  &  50.88$\pm$ 4.83   \\    
\textbf{DBAL}~\cite{gal2017deep}    &     \checkmark                   
& 86.72$\pm$ 4.09    & 88.67$\pm$ 3.06    & 89.04$\pm$ 1.67   & 91.74$\pm$ 0.37  & 45.13$\pm$ 3.90    & 49.58$\pm$ 1.76  &  54.87$\pm$ 0.65   \\ 
\textbf{VAAL}~\cite{sinha2019variational} &  \checkmark                          
& 86.07$\pm$ 3.90    & \textbf{88.95}$\pm$ 2.88    & 90.71$\pm$ 1.49   & 92.11$\pm$ 1.21    & \textbf{46.89}$\pm$ 2.69    & 50.60$\pm$ 0.84  &  57.85$\pm$ 0.56 \\
\textbf{CSAL}~\cite{gao2020consistency}&    \checkmark                         
& \textbf{88.39}$\pm$ 2.51    & 88.49$\pm$ 2.60    & \textbf{91.27}$\pm$ 1.49   & \textbf{92.94}$\pm$ 0.37   & 46.80$\pm$ 4.64    & \textbf{52.83}$\pm$ 1.21  &  \textbf{58.77}$\pm$ 0.62  \\ \midrule[1pt]
\textbf{\method{}}  &         \checkmark                  
& \textbf{89.14}$\pm$ 3.16     & \textbf{91.46}$\pm$ 1.49    & \textbf{93.13}$\pm$ 1.67   & \textbf{95.82}$\pm$ 0.28    & \textbf{48.19} $\pm$ 1.67   & \textbf{57.38}$\pm$ 1.58   & \textbf{61.47}$\pm$ 0.56 \\
\bottomrule[1.5pt]
\end{tabular}
\label{tab:results}
\vspace{-0.2cm}
\end{table*}

\begin{table}
\small
\caption{ \textbf{Experimental results of Macro Precision (MP), Macro F1 score (MF1), Macro Recall (MRC)  on MESCC   dataset.}}
\centering
\setlength{\tabcolsep}{1.4mm}{
\begin{tabular}[width=1\textwidth]{l|ccc|ccc}
\toprule[1.5pt]

\multicolumn{1}{c|}{\multirow{2}{*}{\textbf{Methods}}} & \multicolumn{3}{c|}{MESCC   two-grading}                                                  & \multicolumn{3}{c}{MESCC   six-grading}\\

\cmidrule(l){2-4}\cmidrule(l){5-7}
\multicolumn{1}{c|}{}  & \textbf{MP}  & \textbf{MF1}  & \textbf{MRC}     & \textbf{MP}    & \textbf{MF1}  & \textbf{MRC}     
\\ \cmidrule(l){1-4}\cmidrule(l){5-7} \textbf{P-Labeling}$^{\dagger}$~\cite{lee2013pseudo}  
                                                  &61.15 &62.73	&65.46	&23.13	&22.68&32.20	\\
\textbf{MixMatch}$^{\dagger}$~\cite{berthelot2019mixmatch}  &73.15 &77.20	&84.50	
&28.33	&29.28	&38.81\\
\textbf{FixMatch}$^{\dagger}$~\cite{sohn2020fixmatch} &73.75&77.97	&\textbf{85.60} &	
30.11&31.70	&\textbf{40.24}	\\
\hline
\textbf{R-Labeling}~\cite{figueroa2012active}   &65.86 &66.68	&67.64	&27.58& 27.91	&34.34	\\
\textbf{DBAL}~\cite{gal2017deep}  &76.97 &78.89	&81.24	
&28.13&28.54	&33.61	\\
\textbf{VAAL}~\cite{sinha2019variational}   &75.60 &\textbf{78.91}	&83.79	&31.76 &33.38	&{39.75}	\\

\textbf{CSAL}~\cite{gao2020consistency}  &77.87 &78.87	&79.97	&\textbf{32.55} &\textbf{34.07}	&39.83	\\
\bottomrule[1.5pt]
\textbf{\method{}}  &\textbf{86.09} &\textbf{87.54}	&\textbf{89.15}	&\textbf{36.74 } 	&\textbf{38.77}	&\textbf{45.06}
\\ \bottomrule[1.5pt]
\end{tabular}}
\label{tab:result_1}
\vspace{-0.2cm}
\end{table}

\subsection{Experimental Results}
The overall quantitative results of our framework and baselines on the test set of the MESCC dataset are listed in Table~\ref{tab:results}. From this table, we can observe that our model consistently achieves superiority of performance from 15\% to 30\% labeled data. In particular, under 30\% labeled data, \method{} outperforms  other baselines in terms of accuracy by a large margin of \textbf{\underline{2.88\% $\sim$ 10.21\%}} (two-grading) and \textbf{\underline{2.70\% $\sim$ 18.31\%}} (six-grading) for MESCC medical image classification task.
It is worth noting that almost all AL-based SSL models can improve the model performance by employing an appropriate annotation strategy, except for the R-Labeling. This phenomenon is reasonable as the randomly selected annotation candidates are not informative enough for the SSL model.
In addition, compared to the pure SSL model Fixmatch, \method{} brings a huge reduction in \textbf{\underline{10\% annotation cost (500 samples)} } to reach the similar performance (\method{}: 91.46\%, Fixmatch: 91.09\%). In other words, benefiting from the adversarial unstability selector (AUS) and balanced uncertainty selector (BUS), \method{} can provide more informative samples for the task model with less annotation cost and a significant performance improvement.

We also report the Macro Precision, the Macro F1 score and the Macro Recall of the compared models in Table~\ref{tab:result_1}. Our \method{} once again achieves the best score in terms of all the metrics. Notably, the results of AL-based SSL models fail to maintain consistent superiority compared with the global accuracy in Table~\ref{tab:results}. For instance, the Macro Recall scores of AL-based SSL baselines are inferior to FixMatch. This is because those AL models tend to select annotation candidates with class b0/low-grade, which has the largest data proportion in MESCC dataset. The imbalanced annotation candidates without balanced supervised label information may lead to the misclassification of other class data. Meanwhile, the balanced uncertainty selector in \method{} uniformly chooses the samples with high uncertainty in each predicted class, which helps to maintain remarkable results consistently.

\subsection{In-Depth Analysis}
We further validate several vital issues of the proposed method by answering the four questions as follows. 

$\textbf{Q1: What is the contribution of the individual comp-}$\\
$\textbf{onents  in \method{} to boost the medical image SSL?}$ We conduct an ablation study to illustrate the effectiveness of each component in Table~\ref{tab:ablation}.  Comparing \method{} and \method{}(-AT) (Row 2 vs Row 4), the adaptive threshold significantly contributes 0.84\% and 1.40\% improvement on accuracy. The results of Row 3 and Row 4 show the accuracy improvement of the AUS (unstable sampling) and BUS
(uncertain sampling)
in AL module, respectively. Meanwhile, SSL+RS (Row 1) which combines the SSL module in \method{} and AL random sampling obtains the worst performance. This further validates the superiority of our AL informative selection. Finally, the results indicate that the introduced two AL selection strategies can boost the medical image SSL in a mutually rewarding way.

\begin{table}
\small
\caption{\textbf{Ablation study of different modules in \method{} over MESCC.}   The results are reported when the AL selected samples reaches 20\% data.}
\centering
\setlength{\tabcolsep}{0.5mm}{
\begin{tabular}[width=1\textwidth]{l|cc|cc|cc}
\toprule[1.5pt]

\multicolumn{1}{c|}{\multirow{2}{*}{\textbf{Methods}}} & \multicolumn{2}{c|}{Threshold }                                                  & \multicolumn{2}{c|}{AL Selection} & \multicolumn{2}{c}{ACC } 
\\

\cmidrule(l){2-3}\cmidrule(l){4-5}\cmidrule(l){6-7}
\multicolumn{1}{c|}{}  & \textbf{Adaptive}  & \textbf{Fixed}  & \textbf{Uncertain}     & \textbf{Unstable}    & \textbf{Two}& \textbf{Six}   
\\\cmidrule(l){1-1}\cmidrule(l){2-3}\cmidrule(l){4-5} \cmidrule(l){6-7}
\textbf{SSL+RS}  &\checkmark & 	& 	& &91.45 &53.76	\\ \bottomrule[1.5pt]
\quad\textbf{- AS}  & &\checkmark 	& \checkmark	& \checkmark& 94.98	& 60.07	\\
\quad\textbf{- BUS}   &\checkmark & 	& 	& \checkmark&\textbf{94.06} 	&\textbf{60.44} 	\\
\quad\textbf{- AUS} &\checkmark & 	&\checkmark &	&92.57 &58.03 \\
\toprule[1.5pt]
\textbf{\method{}}  &\checkmark &	&\checkmark	&\checkmark&\textbf{95.82}&	\textbf{61.47}
\\
\toprule[1.5pt]
\end{tabular}}
\vspace{-0.3cm}
\label{tab:ablation}
\end{table}

\begin{figure}[t]
\vspace{-0.1cm}
\includegraphics[width=0.5\textwidth]{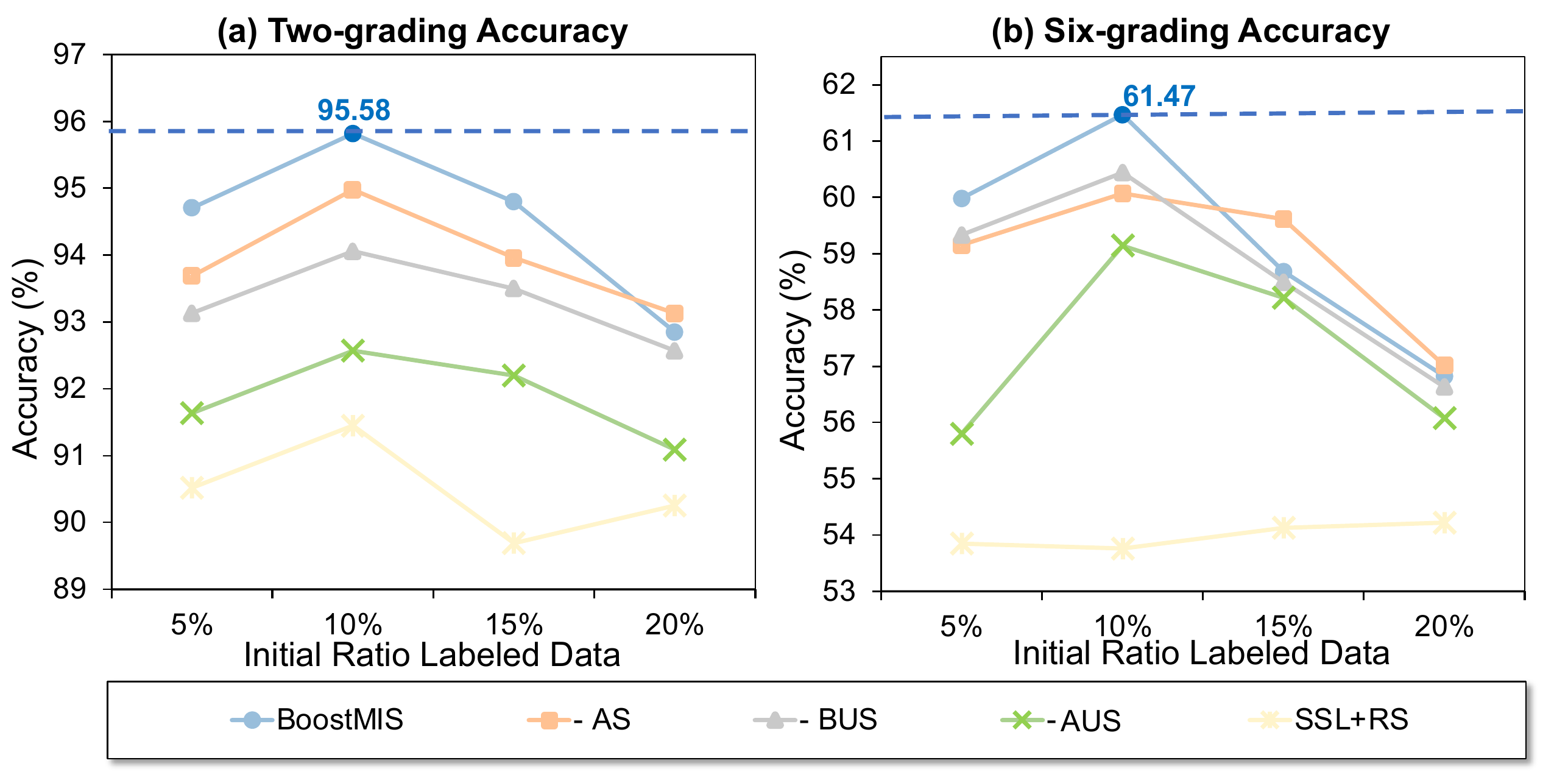}
\centering\caption{ \textbf{Performance comparison under different size of initial labeled data pool $IP$. }}
\vspace{-0.16cm}
\label{initial_pool}
\end{figure}

$\textbf{Q2: How to determine the optimal start size of the in-}$
$\textbf{itial labeled pool?}$
We conduct an exploratory analysis to systematically infer a proper starting size of the initial labeled pool ($IP$) for  \method{}. As shown in Figure~\ref{initial_pool},  using uniform sampling to select different starting sizes, \method{} and its variants achieve different accuracies (optimal start size is 10\% data). For example, the model starting with $IP$=5\% labeled data clearly under-performs the model starting with $IP$ = 10\% labeled samples, when both models reach 30\% labeled samples. We observe that an SSL model with a small start size of labeled data may blindly predict some unlabeled samples into a certain class, due to the fact that the model lacks the data distribution information of these samples. 
This may affect the subsequent AL selection which is based on the prediction from the task model. On the contrary, a big start size may also reduce an AL-based SSL ability (e.g., 10\% and 15\% labeled samples). Although the task model may learn a relatively strong data distribution, the capacity of the AL budget is limited. 
There are not enough AL selected data to improve the robustness of the task model.  
In summary, naively choosing a start size is not advisable, because it may lead to under-utilization of AL-based selection.

\begin{figure}[t]
\includegraphics[width=0.5\textwidth]{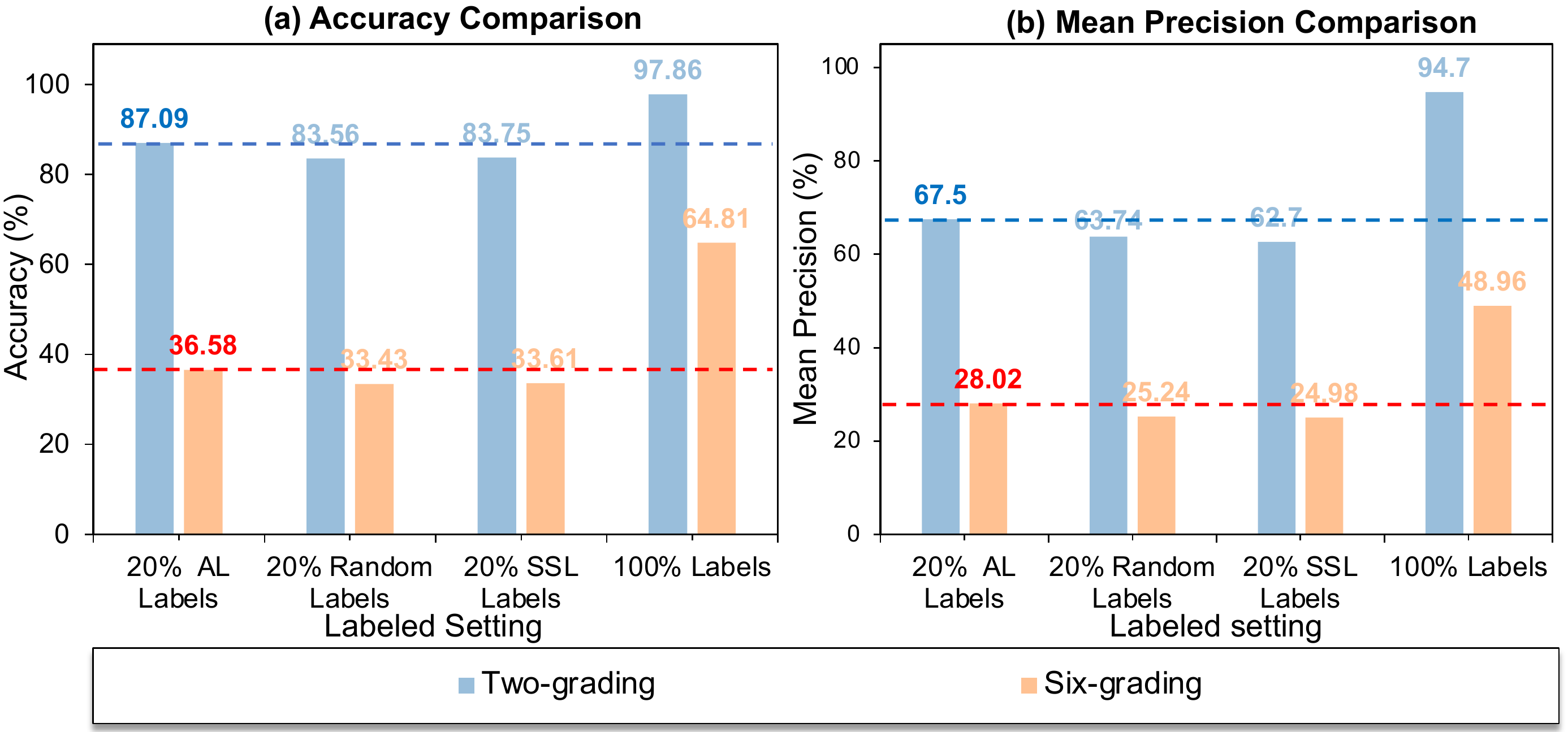}
\centering\caption{\textbf{Performance comparison under supervised learning with different labeled data setting on the MESCC dataset.}}
\vspace{-0.3cm}
\label{al_random}
\end{figure}

$\textbf{Q3: Are these annotation candidates from AL select-}$\\
$\textbf{ion valuably informative?}$ To build insights on the superior performance of our AL selection method, we perform an in-depth analysis of accuracy improvement that benefits from the informative annotation candidates. Figure~\ref{al_random} summarizes the task model's performance using different labeled samples in supervised learning, which indicates the following: First, the performance is similar when using the same scale of randomly selected data and original data with the supervised label. This proves that the original data are not biased and have not hindered the model's learning ability.  Second, training a task model with AL selected samples can represent the global data distribution better. The accuracy results of \textbf{\underline{ 20\% AL samples are competitive}} even comparing the training model with 100\% labeled data.

To further analyze where the improvement comes from,
we  visualize our method’s sample selection behavior via tSNE embedding~\cite{van2008visualizing} in Figure~\ref{tsne_vis}. This visualization result suggests that the AUS tends to select unstable samples near the decision boundary and the BUS can select the representative uncertain samples. These unstable and uncertain samples can smooth the decision boundary for the medical image SSL model and propagate representative label information to unlabeled data, respectively.



\begin{figure}[t]
\includegraphics[width=0.52\textwidth]{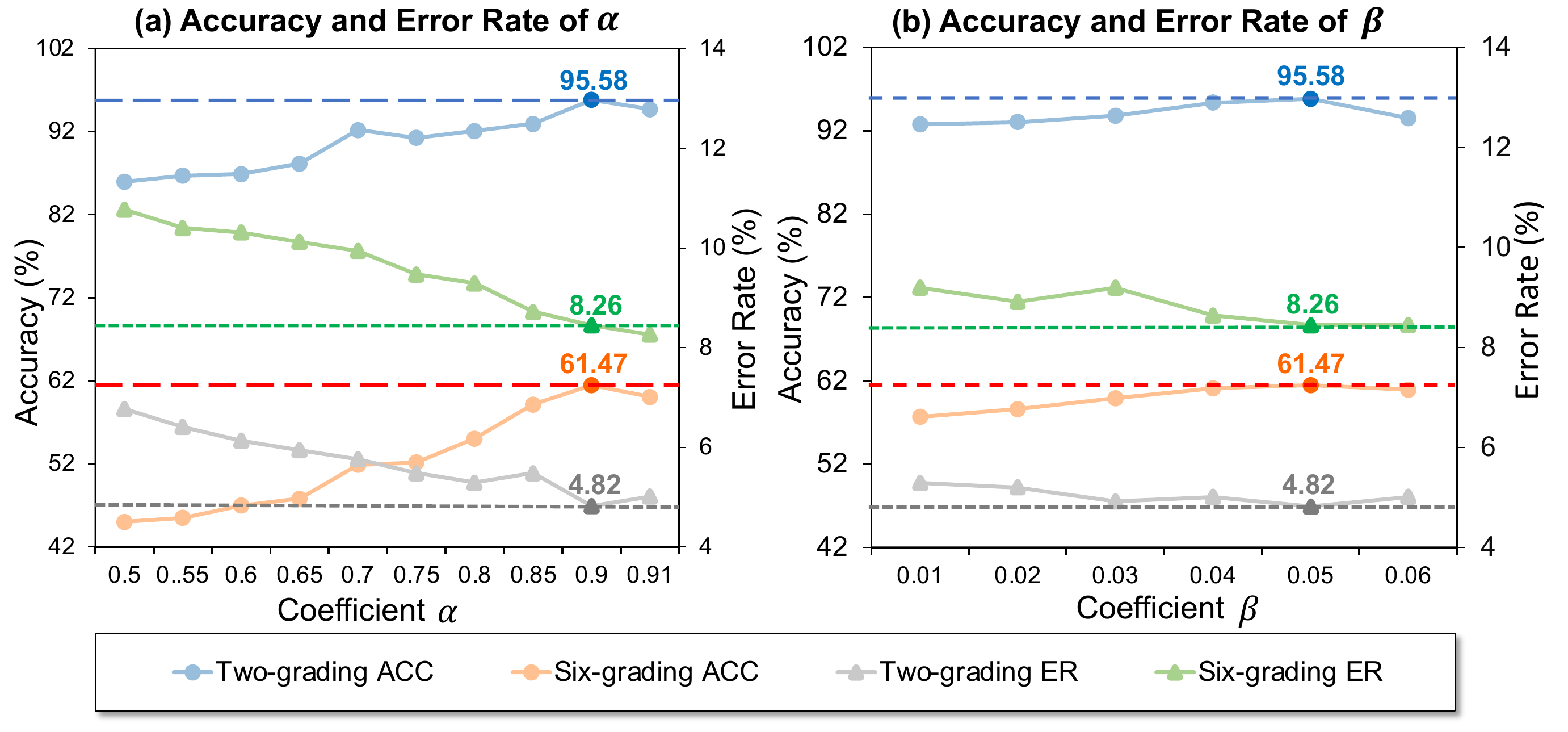}
\centering\caption{\textbf{Ablation studies with respect to different  coefficient $\alpha$ and $\beta$ of adaptive threshold (AS).} }
\label{threshold}
\end{figure}

\begin{figure}[t]
\vspace{-0.2cm}
\includegraphics[width=0.5\textwidth]{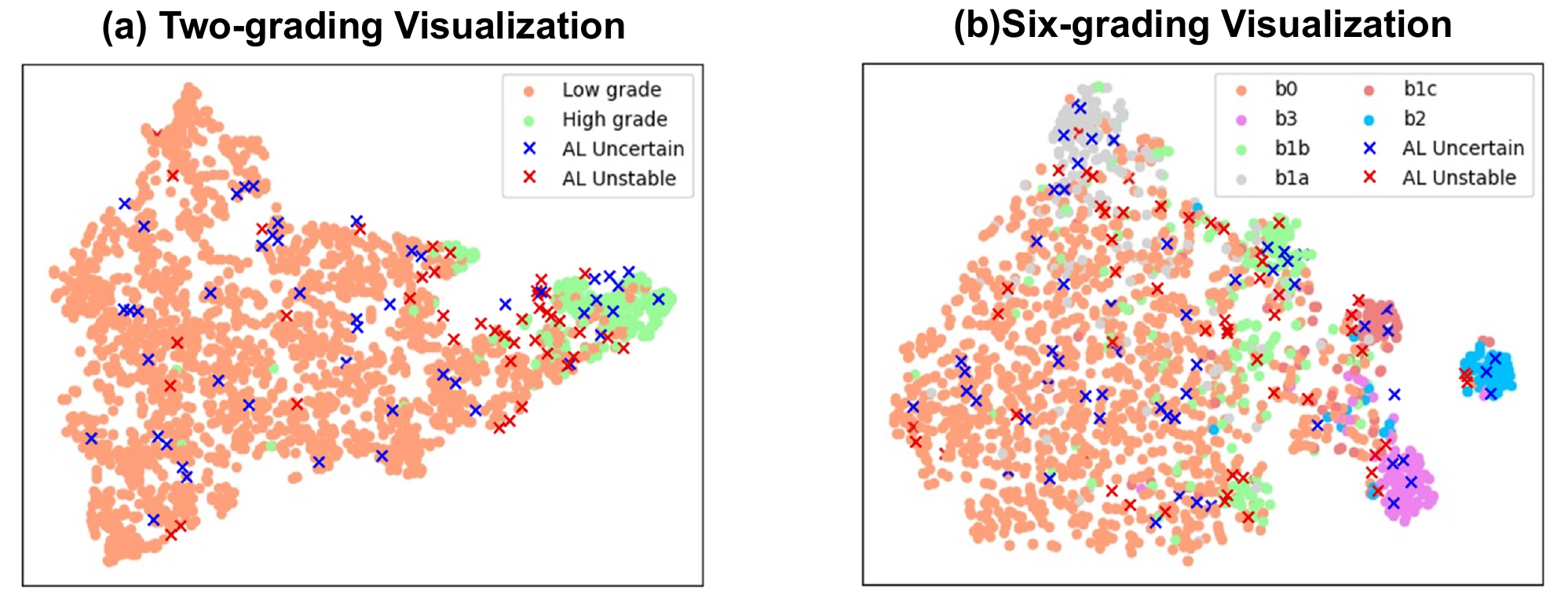}
\centering\caption{  \textbf{The tSNE embeddings of the MESCC  dataset and the AL selection behavior of \method{}. } }
\vspace{-0.1cm}
\label{tsne_vis}
\end{figure}

$\textbf{Q4: Can the model trust the pseudo-labels  based on}$
$\textbf{adaptive threshold?}$
We investigate the impact of hyper-parameters $\alpha$ and $\beta$ of AS for SSL ability. The classification error rate and accuracy of different hyper-parameters on MESCC dataset are shown in Figure~\ref{threshold}. We can observe that the lowest error rate of generated pseudo-labels are 4.83\% and 8.26\% for two-grading and six-grading. This figure also suggests that the optimal choice of $\alpha$ and $\beta$ are around 0.9 and 0.05, respectively, either increasing or decreasing these values results in a performance decay. Moreover, Table~\ref{tab:selected_data} summarizes the number of correct pseudo-labels in the training stage, where our method significantly improves the unlabeled data utilization. Compared with the pseudo labeling SSL method FixMatch, our \method{}  surpasses it by \textbf{\underline{4.31\% and 8.17\%}} relatively on unlabeled data utilization for two-grading and six-grading. This further verifies the superiority of \method{} that can unleash the potential of unlabeled data for better medical image SSL.

\begin{table}[t]
\small
\caption{ \textbf{Number of correct pseudo-labels in SSL.}}
\vspace{-0.1cm}
\centering\setlength{\tabcolsep}{1.2mm}{
\begin{tabular}[width=1\textwidth]{l|cc|cc}
\toprule[1.5pt]
\multicolumn{1}{c|}{\multirow{1}{*}{Methods}} & \multicolumn{1}{c}{Two-grading}&  \multicolumn{1}{c|}{Ratio\%} & \multicolumn{1}{c}{Six-grading} & \multicolumn{1}{c}{Ratio\%}  
\\ \cmidrule(l){1-5} 
\textbf{FixMatch}$^{\dagger}$~\cite{sohn2020fixmatch} &3225 &88.14	&2386 &65.46	\\\cmidrule(l){1-5}
\textbf{\method{}} &\textbf{3370} &\textbf{92.45}	&\textbf{2684} &\textbf{73.63}	
\\ \bottomrule[1.5pt] 
\end{tabular}}
\label{tab:selected_data}
\vspace{-0.2cm}
\end{table}

\section{Conclusion}

In this paper, we propose a medical image semi-supervised learning (SSL) framework \method{}   that takes advantage of the active learning (AL) to unleash the potential of unlabeled data for better medical image analysis.  In the algorithm,  we propose adaptive pseudo-labeling and informative active annotation that leverage the unlabeled medical images and form a closed-loop structure to improve the performance of the medical image SSL model.   The experimental results  show that \method{} significantly outperforms SoTA SSL and AL approaches on medical image classification tasks. 
The proposed framework, which makes use of Apache SINGA~\cite{ooi2015singa} for distributed training, has been integrated into our MLCask\cite{luo2020mlcask} for handling healthcare images and analytics.

\noindent \textbf{Acknowledgement}
We would like to thank the anonymous reviewers for their helpful suggestions and comments. This research is supported by Singapore Ministry of Education Academic Research Fund Tier 3 under MOE’s official grant number MOE2017-T3-1-007, the Singapore Ministry of Health’s National Medical Research Council under its NMRC Clinician-scientist individual research grant number CNIG20nov-0011 and MOH-000725, NCIS Centre Grant Seed Funding Program of Artificial Intelligence for the management of vertebral metastases.


{\small
\bibliographystyle{ieee_fullname}
\bibliography{egbib}
}
\end{document}